\documentclass[twocolumn,showpacs,preprintnumbers,superscriptaddress,amsmath,amssymb]{revtex4}

\usepackage{graphicx}
\usepackage{dcolumn}
\usepackage{bm}
\usepackage{color}
\usepackage{booktabs, longtable}
\usepackage{multirow}
\usepackage{CJK}
\usepackage{ltxtable}
\usepackage{rotating}
\usepackage[colorlinks, citecolor=red]{hyperref}

\begin{document}
\begin{CJK*}{GBK}{song}

\title{Green's function relativistic mean field theory for $\Lambda$ hypernuclei}
\author{Su-Hong Ren}
\author{Ting-Ting Sun}
\email{ttsunphy@zzu.edu.cn}
\author{Wei Zhang}
\affiliation{School of Physics and Engineering, Zhengzhou University, Zhengzhou 450001, China}

\date{\today}

\begin{abstract}
The relativistic mean-field theory with Green's function method is extended to study $\Lambda$ hypernuclei.
Taking hypernucleus $^{61}_{\Lambda}$Ca as an example, the single-particle resonant states
for $\Lambda$ hyperons are investigated by analyzing density of states and the corresponding energies and widths are given. Different behaviors are observed for the resonant states, i.e., the distributions of the very narrow $1f_{5/2}$ and $1f_{7/2}$ states are very similar as bound states while that of the wide $1g_{7/2}$ and $1g_{9/2}$ states are like scattering states.
Besides, the impurity effect of $\Lambda$ hyperons on the single-neutron resonant states are investigated. For most of the resonant states, both the energies and widths decrease with adding more $\Lambda$ hyperons due to the attractive $\Lambda N$ interaction. Finally, the energy level structure of $\Lambda$ hyperons in the Ca hypernucleus isotopes with mass number $A=53-73$ are studied, obvious shell structure and small spin-orbit splitting are found for the single-$\Lambda$ spectrum.
\end{abstract}


\maketitle
\end{CJK*}

\section{Introduction}
\label{Introduction}

Since the first discovery of $\Lambda$ hypernucleus by Danysz
and Pniewski in 1953~\cite{PM1953Danysz_44_348}, the study of hypernuclei has been
attracting great interests of nuclear physicists experimentally~\cite{PRC2001Hotchi_64_044302,PPNP2006Hashimoto_57_564,PTPS2010Nagae_185_299,JP2011Garibaldi_299_012013}.
An important goal of hypernuclear physics is to extract information on the baryon-baryon interactions
including the strangeness of freedom, which are crucial not only for the understanding of hypernuclear structure~\cite{PRL1993Hao_71_1498,NPA1996ZYMa_608_305,PRC2002Tzeng_65_047303,PTPS2010EHiyama_185_1} but also for the study of neutron stars~\cite{PRC2001Hofmann_64_025804,NPA2008Schaffner_804_309,NPA2010Schaffner_835_279,NPA2013Vidana_914_367}.
However, due to the difficulty of
the hyperon-nucleon $(YN)$ and hyperon-hyperon ($YY$) scattering experiments,
there are very limited $YN$ scattering data and no $YY$ scattering data
at all. Thus, in order to shed light on baryon-baryon interactions,
the study of the hypernuclei structure is very important.

The most extensively studied hypernuclear system is the single-$\Lambda$
hypernucleus which consists of one $\Lambda$ hyperon coupled to a nuclear core.
Until now, more than thirty $\Lambda$ hypernuclei ranging from $^{3}_{\Lambda}{\rm H}$
up to $^{208}_{\Lambda}{\rm Pb}$ have been produced experimentally~\cite{PRC2001Hotchi_64_044302,PPNP2006Hashimoto_57_564}.
Several properties of hypernuclei such as the mass number dependence of single-$\Lambda$
binding energy and spin-orbit splitting have been revealed.
Double-$\Lambda$ hypernuclei such as $^{6}_{\Lambda\Lambda}$He~\cite{PRL2001Takahashi_87_212502} have been observed experimentally and
demonstrated the weakly attractive $\Lambda\Lambda$ interaction by the small positive $\Lambda\Lambda$ bond energy.

Being an additional strangeness degree of freedom, a
hyperon is free from nucleon's Pauli exclusion principle, and it may induce many effects
on the nuclear core as an impurity, such as the shrinkage of
the size~\cite{PTP1983TMotoba_70_189,PRC1999EHiyama_59_2351,PRL2010Hiyama_104_212502},
the change of the shape~\cite{PRC2011BNLu_84_014328,PRC2014BNLu_89_044307}, the modification
of its cluster structure~\cite{PRC1996EHiyama_53_2075}, the shift of neutron drip line to a neutron-rich side~\cite{PRC1998Vretenar_57_R1060,EPJA2003Lu_17_19,PRC2008XRZhou_78_054306}, and the occurrence of nucleon and hyperon
skin or halo~\cite{PRC1996EHiyama_53_2075,CPL2002HFLu_19_1775,EPJA2003Lu_17_19}.

Theoretically, many different models have been contributed to study the structure
of $\Lambda$ hypernuclei, such as the cluster model~\cite{PTP1983TMotoba_70_189,SPTP1985TMotoba_81_42,IJMPA1990Bando_05_4021,PRC1999EHiyama_59_2351},
the antisymmetrized molecular dynamics~\cite{PRC2011Isaka_83_044323,PRC2012Isaka_85_034303,PRC2013Isaka_87_021304,
PRC2014Isaka_89_024310},  the shell model~\cite{ AP1971Gal_63_53,AP1978Dalitz_116_167,NPA2008Millener_804_84,NPA2013Millener_914_109},
the mean-field approaches~\cite{AP1976Rayet_102_226,PLB1977RBrockmann_69_167,PLB1981ABouyssy_99_305, NPA1981Rayet_367_381, PTP1988Yamamoto_80_757,PRC1994Mares_49_2472,PTP1994YCSugahara_92_803,PRC1998Vretenar_57_R1060, PRC2007XRZhou_76_034312,PRC2008MTWin_78_054311,PRC2011MWin_83_014301,
PRC2011BNLu_84_014328} and the ab-initio method~\cite{arXiv2014Wirth}.
Among these methods, the mean-field approach has an advantage in that
it can be globally applied from light to heavy hypernuclei.
Recently, both the Skyrme-Hartree-Fock (SHF)~\cite{AP1976Rayet_102_226, NPA1981Rayet_367_381, PTP1988Yamamoto_80_757, PRC2007XRZhou_76_034312,PRC2011MWin_83_014301} and the relativistic mean field
(RMF) model~\cite{PLB1977RBrockmann_69_167, PLB1981ABouyssy_99_305, PRC1994Mares_49_2472,PTP1994YCSugahara_92_803, PRC1998Vretenar_57_R1060,
PRC2008MTWin_78_054311,PRC2011BNLu_84_014328} have been applied to hypernuclear physics.

During the last decades, the RMF model has achieved great successes in ordinary nuclei~\cite{ANP1986Sert_16_1, RPP1989Reinhard_52_439,PPNP1996Ring_37_193, PR2005Vretenar409_101, PPNP2006MengJ_57_470,JPG2015JMeng_42_093101}. In 1977, Brockmann and Weise applied this approach to hypernuclei~\cite{PLB1977RBrockmann_69_167}.
At that time, it had been already observed experimentally that the spin-orbit
splittings in hypernuclei are significantly smaller than that in ordinary nuclei~\cite{PLB1978WBruckner_79_157}.
The relativistic approach is suitable for a discussion of spin-orbit splittings in hypernuclei, as the spin-orbit interaction is naturally emerged with the relativistic framework. It has been applied
to describe single- and multi-$\Lambda$ systems, including
the single-particle (s.p.) spectra of $\Lambda$-hypernuclei and the
spin-orbit interaction, and extended beyond the Lambda
to other strange baryons using SU(3)~\cite{PLB1981Boguta_102_93, NPA1981RBrockmann_355_365,JPG1987MRufa_13_143,ZPA1989Mares_333_209, PLB1990Mares_249_181, PRC1990MRufa_42_2469, NPA1991MChiapparini_529_589,PRC1992JSchaffner_46_322,AP1994JSchaffner_235_35,PRC1994Mares_49_2472,NPA1996ZYMa_608_305,PRC1998Vretenar_57_R1060, PRC2016TTSun_94_064319}.

Hyperon halo may occur with the rapid development of radioactive ion beam facilities. For the halo structure, continuum and
resonant states play crucial role, especially those with low orbital angular momenta $l$~\cite{PRC1996DobaczewskiJ_53_2809,PRL1996Meng_77_3963, PRC2003Sandulescu_68_054323,PPNP2006MengJ_57_470}.
For example, in ordinary nuclei, further studies have shown that the s.p.~resonant states are key factors to many exotic nuclear phenomena, such as the halo~\cite{PRL1996Meng_77_3963,PRL1997Poschl_79_3841}, giant halo~\cite{PRL1998MengJ_80_460,PRC2002MengJ_65_041302,PRC2003Sandulescu_68_054323,SCP2003ZhangSQ_46_632, PRC2006TerasakiJ_74_054318,PRC2006Grasso_74_064317}, and deformed halo~\cite{PRC2010Hamamoto_81_021304,PRC2010Zhou_82_011301}. To study the s.p.~resonant states, many techniques have been developed based on the conventional scattering theory~\cite{PR1947Wigner_72_29,Book1972Taylor-ScatteringTheor, PRL1987Hale_59_763, PRC1991Humblet_44_2530,PRC2002CaoLG_66_024311,PRC2010LiZP_81_034311,PRL2012LuBN_109_072501, PRC2013LuBN_88_024323} or the bound-state-like methods~\cite{PRA1970Hazi_1_1109,PR1983Ho_99_1,Book1989Kukulin-TheorResonance}.
Meanwhile, the combinations of a number of the techniques for the s.p.~resonant states with the RMF theory have been developed. For examples, the RMF theory with the $S$-matrix (RMF-S)~\cite{PRC2002CaoLG_66_024311}; the RMF theory with the analytic continuation in the coupling constant approach (RMF-ACCC)~\cite{CPL2001YangSC_18_196,PRC2004ZhangSS_70_034308,PRC2005GuoJY_72_054319,PRC2006GuoJY_74_024320}; the RMF theory with the real stabilization method (RMF-RSM)~\cite{PRC2008ZhangL_77_014312}; the RMF theory with complex scaling method (RMF-CSM)~\cite{PRC2010GuoJY_82_034318}, and the
RMF theory with Green's function method (RMF-GF)~\cite{PRC2014TTSun_90_054321,JPG2016TTSun_43_045107}.

Green's function method~\cite{SJNP1987Belyaev_45_783,Book2006Eleftherios-GF} has been demonstrated to be an efficient tool for describing the s.p.~resonant states~\cite{PRC2014TTSun_90_054321,JPG2016TTSun_43_045107}. It has been widely applied in nuclear physics to properly take into account the continuum, e.g., the ground state studies based on (non-relativistic) Hartree-Fock-Bogoliubov (HFB) theory~\cite{PRC2009Oba_80_024301,PRC2011ZhangY_83_054301,PRC2012ZhangY_86_054318, arXivSun2014}, and the excited state studies based on the quasiparticle random-phase-approximation (QRPA) theory~\cite{NPA2001Matsuo696_371,PTPS2002Matsuo_146_110} and the relativistic continuum random-phase-approximation (RCRPA) theory~\cite{PRC2009DaoutidisRing80_024309, PRC2010DingY_82_054305}. It is found that the Green's function method has the following advantages: (a) treating the discrete bound states and the continuum on the same footing, (b) giving both the energies and widths of the resonant states directly, and (c) taking into account the correct asymptotic behaviors for the wave functions.

In this paper, we extend the RMF-GF model to include the $\Lambda$ hyperon in coordinate space, detailed formula of $\Lambda N$ interaction and construction of Green's function for $\Lambda$ hyperons are presented. We apply this newly developed theory to three cases.
First, taking $^{61}_{\Lambda}$Ca as an example, we apply the RMF-GF model to study the single-$\Lambda$ resonant states. By analyzing the density of states, the s.p.~energy for bound states and energies and widths for the resonant states are given. Second, taking $^{60}$Ca, $^{61}_{\Lambda}$Ca and $^{62}_{2\Lambda}$Ca as examples, we investigate the impurity effect of $\Lambda$ particle and focus on the influences of $\Lambda$ hyperons on the single-neutron resonant states. Third, the s.p.~level for $\Lambda$ hyperon in the Ca hypernucleus isotopes are given and the shell structure and spin-orbit splitting are discussed.

The paper is organized as follows. In Sec.~\ref{Theory}, we present the formalism of RMF-GF model for $\Lambda$-hypernuclei.
After the numerical details in Sec.~\ref{Sec:Numerical}, we present the results and discussions in Sec.~\ref{Sec:Results}. Finally a summary is drawn in Sec.~\ref{Sec:summary}.

\section{Theoretical Framework}\label{Theory}
\subsection{RMF model for $\Lambda$ hypernuclei}

The starting point of the meson-exchange RMF
model for $\Lambda$ hypernuclei is a covariant Lagrangian density
\begin{equation}
 \mathcal{L} = \mathcal{L}_N + \mathcal{L}_{\Lambda} \:,%
 \label{Eq:lag}
\end{equation}
where $\mathcal{L}_N$ is the standard RMF Lagrangian density for nucleons~\cite{PPNP1996Ring_37_193,PPNP2001Mueller_46_359,PPNP2006MengJ_57_470,JPG2015JMeng_42_093101}, and
$\mathcal{L}_{\Lambda}$ is the Lagrangian density for $\Lambda$ hyperons~\cite{PRC1994Mares_49_2472}. Since the $\Lambda$ hyperon is charge neutral with isospin $0$, only the couplings with $\sigma$- and $\omega$-mesons are included,
\begin{eqnarray}
 && \mathcal{L}_{\Lambda} = \bar{\psi}_{\Lambda} \Big[
 i\gamma^\mu\partial_\mu - m_\Lambda -
 g_{\sigma\Lambda}\sigma - g_{\omega\Lambda}\gamma^\mu\omega_\mu
\\
 &&\hskip6.5mm
 - \frac{f_{\omega\Lambda}}{2m_\Lambda}
 \sigma^{\mu\nu}\partial_\nu\omega_\mu
 \Big]\psi_\Lambda \:,
\nonumber
\label{EQ:Lag-LN}
\end{eqnarray}
where $m_\Lambda$
is the mass of the $\Lambda$ hyperon,
$g_{\sigma\Lambda}$ and $g_{\omega\Lambda}$
are the coupling constants with the
$\sigma$- and $\omega$-mesons, respectively.
The last term in $\mathcal{L}_\Lambda$ is the tensor coupling
with the $\omega$ field~\cite{PLB1980Noble_89_325}, which is
related with the s.p.~spin-orbit splitting of $\Lambda$ hyperons.

For a system with time-reversal symmetry, the space-like components of the vector $\omega_\mu$ field vanish,
only leaving the time components $\omega_0$. With the mean-field and no-sea approximations, the s.p.~Dirac equations for baryons
and the Klein-Gordon equations for mesons and photon can be obtained by the variational procedure.

The Dirac equation for $\Lambda$ hyperon is
\begin{eqnarray}
[\bm{\alpha\cdot p}+\beta(m_{\Lambda}+S(\bm r))+V(\bm r)+T({\bm r})]\psi_{i,\Lambda}(\bm r)&&\nonumber\\
=\varepsilon_i\psi_{i,\Lambda}(\bm r),&&
\label{EQ:Dirac}
\end{eqnarray}
where $\bm{\alpha}$ and $\beta$ are the Dirac matrices, $S({\bm r})$, $V({\bm r})$ and $T({\bm r})$ are the scalar, vector and tensor potentials for $\Lambda$ hyperons, respectively, and
\begin{subequations}
\begin{eqnarray}
 &&S=g_{\sigma\Lambda}\sigma,\\
 &&V= g_{\omega\Lambda}\omega_0,\\
 &&T=-\frac{f_{\omega\Lambda}}{2m_{\Lambda}}i{\bm \gamma}\cdot{\bm \nabla}\omega_0,
\end{eqnarray}%
 \label{EQ:potential}%
\end{subequations}%
with the ${\bm \gamma}$ matrix, i.e., $\gamma^k=\left(
                                               \begin{array}{cc}
                                                 0 & \sigma^k \\
                                                 -\sigma^k & 0 \\
                                               \end{array}
                                             \right)
$ where $k$ runs from $1$ to $3$ and $\sigma^k$ are Pauli matrices.

The Klein-Gordon equations for the $\sigma$- and $\omega$-mesons are changed to
\begin{equation}
(-\Delta+m_{\phi}^{2})\phi=S_{\phi},
\label{EQ:KG}
\end{equation}
with the source terms
\begin{equation}
S_{\phi}=
\left\{
  \begin{array}{ll}
    -g_{\sigma}\rho_S-g_{\sigma\Lambda}\rho_{S\Lambda}-g_{2}\sigma^2-g_{3}\sigma^3 & \hbox{for $\sigma$;} \\
    {\displaystyle g_{\omega}\rho_V+g_{\omega\Lambda}\rho_{V\Lambda}+\frac{f_{\omega\Lambda}}{2m_{\Lambda}}\partial_{k}{\bm j}_{T\Lambda}^{0k}
        -c_{3}\omega_{0}^{3} }& \hbox{for $\omega$,}
  \end{array}
\right.
\end{equation}
where $m_{\phi}(\phi=\sigma,\omega)$ are the corresponding meson masses, $g_{\sigma}$, $g_{\omega}$, $g_{2}$, $g_{3}$, and $c_{3}$ are the parameters for the nucleon-nucleon ($NN$) interaction in the Lagrangian density $\mathcal{L}_{N}$, $\rho_{S}(\rho_{S\Lambda})$, $\rho_{V}(\rho_{V\Lambda})$ are the scalar and baryon densities for the nucleons(hyperons), respectively, and ${\bm j}_{T\Lambda}^{0}$ is the tensor density for $\Lambda$ hyperons.

With the upper $G_{i,\Lambda}(\bm r)$ and lower $F_{i,\Lambda}({\bm r})$ components of Dirac spinor $\psi_{i,\Lambda}({\bm r})$, the densities for $\Lambda$ hyperons can be expressed as
\begin{subequations}
\begin{eqnarray}
&&\rho_{S\Lambda}({\bm r})=\sum_{i=1}^{A_\Lambda}[G_{i,\Lambda}(\bm r)G_{i,\Lambda}^*(\bm r)-F_{i,\Lambda}(\bm r)F_{i,\Lambda}^*(\bm r)],\\
&&\rho_{V\Lambda}({\bm r})=\sum_{i=1}^{A_\Lambda}[G_{i,\Lambda}(\bm r)G_{i,\Lambda}^*(\bm r)+F_{i,\Lambda}(\bm r)F_{i,\Lambda}^*(\bm r)],\\
&&{\bm j}_{T\Lambda}^{0}(\bm r)=\sum_{i=1}^{A_\Lambda}[G_{i,\Lambda}(\bm r)F_{i,\Lambda}^*(\bm r)+F_{i,\Lambda}(\bm r)G_{i,\Lambda}^*(\bm r)]{\bm n},~~~~~~~
\end{eqnarray}%
\label{EQ:density-GF}%
\end{subequations}%
where ${\bm n}$ is the angular unit vector. The number of $\Lambda$ hyperons $A_{\Lambda}$ is calculated by the integral of the
hyperon density $\rho_{V\Lambda}({\bm r})$ in the coordinate space as
\begin{equation}
A_{\Lambda}=\int d^3 r\rho_{V\Lambda}({\bm r}).
\label{EQ:PNumber}
\end{equation}
And the total baryon (mass) number $A$ in hypernuclei is the summation of the neutron, proton and $\Lambda$ hyperon particle numbers.

The Dirac equation for nucleons and Klein-Gordon equations for $\rho$ meson and photon are the same as those in the standard RMF model. All these coupled equations together with Eqs.~(\ref{EQ:Dirac})-(\ref{EQ:PNumber}) are solved by iteration in the coordinate space.

\subsection{Green's function method}

A Green's function $\mathcal{G}(\bm r,\bm r'; \varepsilon)$ describes the propagation of a particle with an energy $\varepsilon$ from coordinate $\bm r$ to $\bm r'$. In the RMF-GF theory~\cite{PRC2014TTSun_90_054321,JPG2016TTSun_43_045107}, the Green's function method is taken to solve the Dirac equation in coordinate space and the relativistic s.p.~Green's function obeys
\begin{equation}
[\varepsilon-\hat{h}(\bm r)]\mathcal{G}(\bm r,\bm r';\varepsilon)=\delta(\bm r-\bm r'),%
 \end{equation}%
where $\hat{h}(\bm r)$ is the Dirac Hamiltonian, and energy $\varepsilon$ can be any value on a energy complex plane. For $\Lambda$ hyperons, $\hat{h}(\bm r)=\bm{\alpha\cdot p}+\beta(m_{\Lambda}+S(\bm r))+V(\bm r)+T({\bm r})$. With a complete set of eigenstates~$\psi_{i,\Lambda}(\bm r)$~and eigenvalues~$\varepsilon_i$, the Green's function for $\Lambda$ hyperons can be represented as
\begin{equation}
 \mathcal{G}(\bm r,\bm r';\varepsilon)= \sum_i\frac{\psi_{i,\Lambda}(\bm r)\psi_{i,\Lambda}^\dagger(\bm r')}{\varepsilon-\varepsilon_i},
 \label{EQ:GF}
\end{equation}
where~$\Sigma_i$~is summation for the discrete states and integral for the continuum explicitly. Green's function in Eq.~(\ref{EQ:GF}) is analytic
on the complex energy plane with the poles at eigenvalues~$\varepsilon_i$. Corresponding to the upper $G_{i,\Lambda}({\bm r})$ and lower $F_{i,\Lambda}({\bm r})$ components of the Dirac spinor~$\psi_{i,\Lambda}({\bm r})$, the Green's function for the Dirac equation is in a form of a~$2\times2$~matrix,
 \begin{equation}
\mathcal{G}(\bm r,\bm r';\varepsilon)=
\left(
        \begin{array}{cc}
          \mathcal{G}^{(11)}(\bm r,\bm r';\varepsilon)&  \mathcal{G}^{(12)}(\bm r,\bm r';\varepsilon) \\
          \mathcal{G}^{(21)}(\bm r,\bm r';\varepsilon) &  \mathcal{G}^{(22)}(\bm r,\bm r';\varepsilon)\\
        \end{array}
      \right).
\end{equation}

\begin{figure}[t!]
 \includegraphics[width=0.45\textwidth]{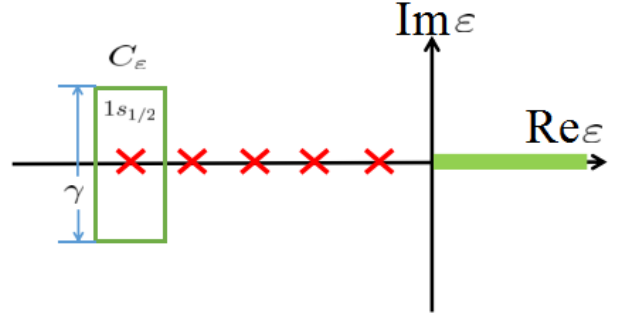}
 \caption{(Color online)  Contour path $C_{\varepsilon}$ to perform the integrals of the Green's function on the complex energy plane. The path is chosen to be a rectangle with height $\gamma$ and enclose only the $1s_{1/2}$ orbit. The red crosses denote the discrete single-$\Lambda$ states and the green thick line denotes the continuum.}
 \label{Fig1}
\end{figure}

According to Cauchy's theorem, the nonlocal scalar density $\rho_{S\Lambda}(\bm r,\bm r')$, vector density $\rho_{V\Lambda}(\bm r,\bm r')$ and tensor density ${\bm j}_{T\Lambda}^0({\bm r},{\bm r'})$ for $\Lambda$ hyperons can be calculated
by the integrals of the Green's function on the complex energy plane,
\begin{subequations}
\begin{eqnarray}
&&\!\!\!\!\!\rho_{S\Lambda}({\bm r},{\bm r'})=\sum_{i=1}^{A_\Lambda}[G_{i,\Lambda}(\bm r)G_{i,\Lambda}^*(\bm r')\!-\!F_{i,\Lambda}(\bm r)F_{i,\Lambda}^*(\bm r')]~~~~~~~~\\
&&~~~~~~~~~~~=\frac{1}{2\pi i}\oint_{C_{\varepsilon}} d\varepsilon[ \mathcal{G}^{(11)}(\bm r,\bm r';\varepsilon)\!-\!\mathcal{G}^{(22)}(\bm r,\bm r';\varepsilon)],\nonumber
\label{EQ:rhos}
\end{eqnarray}
\begin{eqnarray}
&&\!\!\!\!\!\rho_{V\Lambda}({\bm r},{\bm r'})=\sum_{i=1}^{A_\Lambda}[G_{i,\Lambda}(\bm r)G_{i,\Lambda}^*(\bm r')\!+\!F_{i,\Lambda}(\bm r)F_{i,\Lambda}^*(\bm r')]~~~~~~~~\\
&&~~~~~~~~~~~=\frac{1}{2\pi i}\oint_{C_{\varepsilon}} d\varepsilon[ \mathcal{G}^{(11)}(\bm r,\bm r';\varepsilon)\!+\!\mathcal{G}^{(22)}(\bm r,\bm r';\varepsilon)],\nonumber
\label{EQ:rhov}
\end{eqnarray}
\begin{eqnarray}
&&\!\!\!\!\!{\bm j}_{T\Lambda}^{0}({\bm r},{\bm r'})=\sum_{i=1}^{A_\Lambda}[G_{i,\Lambda}(\bm r)F_{i,\Lambda}^*(\bm r')\!+\!F_{i,\Lambda}(\bm r)G_{i,\Lambda}^*(\bm r')]{\bm n}~~~~~~~\\
&&~~~~~~~~~~~=\frac{1}{2\pi i}\oint_{C_{\varepsilon}} d\varepsilon[ \mathcal{G}^{(12)}(\bm r,\bm r';\varepsilon)\!+\!\mathcal{G}^{(21)}(\bm r,\bm r';\varepsilon)]{\bm n},\nonumber
\label{EQ:rho-tensor}%
\end{eqnarray}%
\label{EQ:density-GF}%
\end{subequations}%
where $C_{\varepsilon}$ is the contour path for the integral of Green's function on the complex energy plane
shown in Fig.~\ref{Fig1}.

With the spherical symmetry, Green's function and densities can be expanded as radial and angular parts,
\begin{subequations}
\begin{eqnarray}
&&\mathcal{G}(\bm r,\bm r';\varepsilon)=\sum_{\kappa m}Y^{l}_{jm}(\theta,\phi)\frac{\mathcal{G}_{\kappa}(r,r';\varepsilon)}{rr'}Y^{l*}_{jm}(\theta',\phi'),~~~~~~~~\\
&&\rho_{S\Lambda}({\bm r},{\bm r'})=\sum_{\kappa m}Y^{l}_{jm}(\theta,\phi)\rho_{S\Lambda,\kappa}({r},{r'})Y^{l*}_{jm}(\theta',\phi'),\\
&&\rho_{V\Lambda}({\bm r},{\bm r'})=\sum_{\kappa m}Y^{l}_{jm}(\theta,\phi)\rho_{V\Lambda,\kappa}({r},{r'})Y^{l*}_{jm}(\theta',\phi'),\\
&&{\bm j}_{T\Lambda}^{0}({\bm r},{\bm r'})=\!\sum_{\kappa m}Y^{l}_{jm}(\theta,\phi){\bm j}_{T\Lambda,\kappa}^{0}({r},{r'})Y^{l*}_{jm}(\theta',\phi').
\end{eqnarray}
\end{subequations}
And they are only decided by the radial part, which is characterized with the quantum number
$\kappa=(-1)^{j+l+1/2}(j+1/2)$ and orbits with the same $\kappa$ is defined as a ``block". The radial parts of the
local scalar density $\rho_{S\Lambda}(\bm r)=\rho_{S\Lambda}(\bm r,\bm r)$, vector density $\rho_{V\Lambda}(\bm r)=\rho_{V\Lambda}(\bm r,\bm r)$ and tensor density
${\bm j}_{T\Lambda}^0({\bm r})={\bm j}_{T\Lambda}^0({\bm r},{\bm r})$ can be expressed by the radial part of Green's function as
\begin{subequations}
\begin{eqnarray}
\rho_{S\Lambda}(r)&=&\frac{1}{4\pi r^2}\frac{1}{2\pi i}\sum_{\kappa}(2j+1)
\label{Eq:Rrhos}\\
&&\times\oint_{C_\varepsilon} d\varepsilon[\mathcal{G}_{\kappa}^{(11)}( r,r;\varepsilon)-\mathcal{G}_{\kappa}^{(22)}(r,r;\varepsilon)],\nonumber\\
\rho_{V\Lambda}(r)&=&\frac{1}{4\pi r^2}\frac{1}{2\pi i}\sum_{\kappa}(2j+1)
\label{Eq:Rrhov}\\
&&\times\oint_{C_{\varepsilon}} d\varepsilon[\mathcal{G}_{\kappa}^{(11)}( r,r;\varepsilon)+\mathcal{G}_{\kappa}^{(22)}(r,r;\varepsilon)],\nonumber\\
{\bm j}_{T\Lambda}^{0}(r)&=&\frac{1}{4\pi r^2}\frac{1}{2\pi i}\sum_{\kappa}(2j+1)
\label{Eq:Rrhot}\\
&&\times\oint_{C_{\varepsilon}} d\varepsilon[\mathcal{G}_{\kappa}^{(12)}( r,r;\varepsilon)+\mathcal{G}_{\kappa}^{(21)}(r,r;\varepsilon)]{\bm n},\nonumber
\end{eqnarray}%
\label{EQ:density-R-GF}%
\end{subequations}%
Note that for the $\Lambda$ hyperons occupying the $1s_{1/2}$ orbit, the degeneracy is $2j+1=2$. It is half occupied for single-$\Lambda$ hypernuclei and fully occupied for double-$\Lambda$ hypernuclei.

Different from the standard RMF model, in the RMF-GF model, from the densities given by the Green's function~(\ref{EQ:density-R-GF}), one can solve the Klein-Gordon Eq.~(\ref{EQ:KG}) to obtain the $\sigma$- and $\omega$-fields, and then calculate the single-$\Lambda$ potentials $V(\bm r)$, $S(\bm r)$ and $T({\bm r})$ in Eq.~(\ref{EQ:potential}), and the Dirac equation is solved again to provide new Green's functions. In this way, the RMF coupled equations can be solved by iteration self-consistently.

In the RMF-GF theory, the energies of the s.p.~bound states as well as the energies and widths of the s.p.~resonant states can be extracted from the density of states $n(\varepsilon)$~\cite{PRC2014TTSun_90_054321,JPG2016TTSun_43_045107},
\begin{equation}
n(\varepsilon)=\sum_{i}\delta(\varepsilon-\varepsilon_{i}),
\label{EQ:DOS}
\end{equation}
where $\varepsilon_{i}$ is the eigenvalue of the Dirac equation, $\varepsilon$ is a real s.p.~energy, $\sum_{i}$ includes the summation for the discrete states and the integral for the continuum, and $n(\varepsilon)d\varepsilon$ gives the number of states in the interval $[\varepsilon, \varepsilon+d\varepsilon]$. For the bound states, the density of states $n(\varepsilon)$ exhibits discrete $\delta$-function at $\varepsilon=\varepsilon_{i}$, while in the continuum $n(\varepsilon)$ has a continuous distribution.

In the spherical case, Eq.~(\ref{EQ:DOS}) becomes
\begin{equation}
n(\varepsilon)=\sum_{\kappa}n_{\kappa}(\varepsilon),
\label{EQ:RDOS}
\end{equation}
where $n_{\kappa}(\varepsilon)$ is the density of states for a block characterized by the quantum number $\kappa$.
By introducing an infinitesimal imaginary part $``i\epsilon"$ to energy $\varepsilon$, it can be proved that the density of states can be obtained by integrating the imaginary part of the Green's function over the coordinate space, and in the spherical case, it is~\cite{PRC2014TTSun_90_054321}
\begin{eqnarray}
n_{\kappa}(\varepsilon)&=&-\frac{2j+1}{\pi }\int d{r}{\rm Im}[\mathcal{G}_{\kappa}^{(11)}({r},{r};\varepsilon+i\epsilon)\nonumber \\
~&&+\mathcal{G}_{\kappa}^{(22)}({r},{ r};\varepsilon+i\epsilon)].
\label{EQ:DOS-lj}
\end{eqnarray}
Moreover, with this infinitesimal imaginary part $``i\epsilon"$, the density of states for discrete s.p.~states in shape of $\delta$-function~(no width) is simulated by a Lorentzian function with the full-width at half-maximum (FWHM) of $2\epsilon$.

\subsection{Construction of Green's function}

In the spherical case, for a given single-$\Lambda$ energy $\varepsilon$ and quantum number $\kappa$, the Green's function $\mathcal{G}_{\kappa}(r,r';\varepsilon)$ for the radial form of Dirac Eq.~(\ref{EQ:Dirac}) can be constructed as~\cite{Book2006Eleftherios-GF, PRC2009DaoutidisRing80_024309, PRC2010DingY_82_054305, PRC2014TTSun_90_054321,JPG2016TTSun_43_045107}
\begin{eqnarray}
&&\mathcal{G}_{\kappa}(r,r';\varepsilon)=\frac{1}{W_{\kappa}(\varepsilon)}
\left[\theta(r-r')
\phi_{\kappa}^{(2)}(r,\varepsilon)\phi_{\kappa}^{(1)\dag}(r',\varepsilon)\right.\nonumber\\
&&~~~~~~~~~~~~~~~ \left.+\theta(r'-r)\phi_{\kappa}^{(1)}(r,\varepsilon)\phi_{\kappa}^{(2)\dag}(r';\varepsilon)
\right],
\label{EQ:GF-RDirac}
\end{eqnarray}
where $\theta(r-r')$ is the radial step function, $\phi_{\kappa}^{(1)}(r,\varepsilon)$ and $\phi_{\kappa}^{(2)}(r,\varepsilon)$ are two linearly independent Dirac spinors for $\Lambda$ hyperons,
\begin{equation}
\phi_{\kappa}^{(1)}(r,\varepsilon)=\left(
                                     \begin{array}{c}
                                       G_{\kappa}^{(1)}(r,\varepsilon) \\
                                       F_{\kappa}^{(1)}(r,\varepsilon) \\
                                     \end{array}
                                   \right),~~~
                                   \phi_{\kappa}^{(2)}(r,\varepsilon)=\left(
                                     \begin{array}{c}
                                       G_{\kappa}^{(2)}(r,\varepsilon) \\
                                       F_{\kappa}^{(2)}(r,\varepsilon) \\
                                     \end{array}
                                   \right),
                                   \label{EQ:phi1_phi2}
\end{equation}
and $W_{\kappa}(\varepsilon)$ is the Wronskian function defined by
\begin{equation}
W_{\kappa}(\varepsilon)=G_{\kappa}^{(1)}(r,\varepsilon)F_{\kappa}^{(2)}(r,\varepsilon)-G_{\kappa}^{(2)}(r,\varepsilon)F_{\kappa}^{(1)}(r,\varepsilon).
\label{EQ:Wronskian}
\end{equation}
and it is independent with coordinate $r$, i.e., $dW_{\kappa}(\varepsilon)/dr=0$.

The Dirac spinor $\phi^{(1)}_{\kappa}(r)$ is regular at the origin and $\phi^{(2)}_{\kappa}(r)$ at $r\rightarrow\infty$ is oscillating outgoing for $\varepsilon >0$ and exponentially decaying for $\varepsilon <0$. Explicitly, Dirac spinor $\phi^{(1)}_{\kappa}(r,\varepsilon)$
 at $r\rightarrow0$ satisfies
\begin{eqnarray}
 \phi^{(1)}_{\kappa}(r,\varepsilon) &\longrightarrow& r\left(
                                                       \begin{array}{c}
                                                         j_l(k r) \\
                                                         \frac{\kappa}{|\kappa|}\frac{\varepsilon-V-S}{k}j_{\tilde{l}}(kr)\\
                                                       \end{array}
                                                     \right),\nonumber\\
                                   &\longrightarrow&\left(
                                                    \begin{array}{c}
                                                      \frac{r}{(2l+1)!!}(kr)^l \\
                                                      \frac{\kappa}{|\kappa|}
                                                      \frac{r(\varepsilon-V-S)}{k(2\tilde{l}+1)!!}(k r)^{\tilde{l}} \\
                                                    \end{array}
                                                  \right),
                                                  \label{Eq:behavior_r0}
 \end{eqnarray}
where $k^2=(\varepsilon-V-S)(\varepsilon-V-S+2m_{\Lambda})>0$, quantum
number $\tilde{l}$ is defined as $\tilde{l}=l+(-1)^{j+l+1/2}$, and $j_l(k r)$ is the spherical Bessel function of the first kind.

\begin{figure}[t!]
 \includegraphics[width=0.45\textwidth]{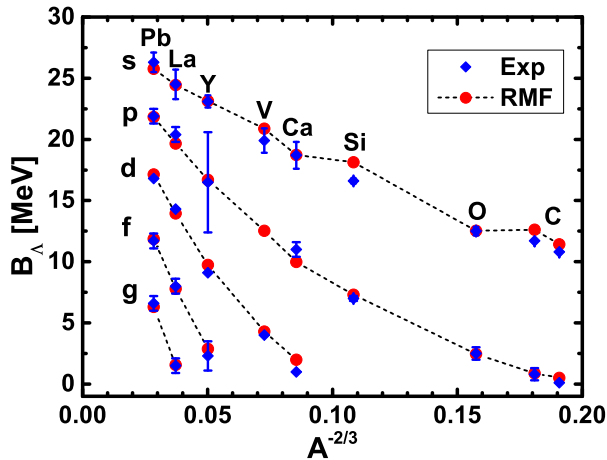}
 \caption{(Color online) Single-$\Lambda$ binding energies $B_{\Lambda}$ for the $\Lambda$-hypernuclei from $^{12}_{\Lambda}$C to $^{208}_{\Lambda}$Pb calculated with RMF-GF method and compared with the experimental data~\cite{PRC2001Hotchi_64_044302,PPNP2006Hashimoto_57_564}.}
 \label{Fig2-added}
\end{figure}

\begin{figure*}[t]
 \includegraphics[width=0.8\textwidth]{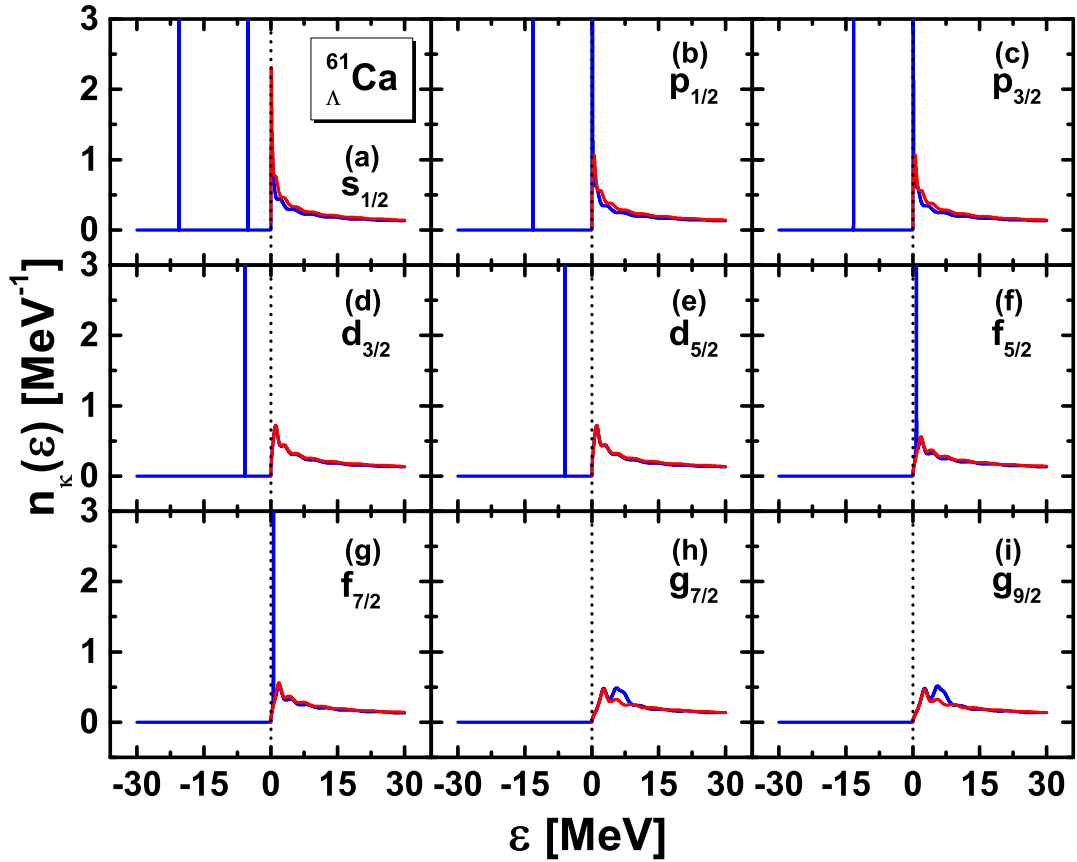}
 \caption{(Color online) Density of states $n_{\kappa}(\varepsilon)$ of $\Lambda$ hyperon for different blocks $\kappa$ in $^{61}_{\Lambda}{\rm Ca}$ calculated with RMF-GF method (blue solid line) and compared with $n_{\kappa}(\varepsilon)$ for free particles obtained with potentials $V=S=0$ (red solid line). The dotted line in each panel indicates the position of the continuum threshold.}
 \label{Fig2}
\end{figure*}

The Dirac spinor $\phi^{(2)}_{\kappa}(r,\varepsilon)$ at $r\rightarrow \infty$ satisfies
 \begin{eqnarray}
 \phi^{(2)}_{\kappa}(r,\varepsilon)&\longrightarrow&\left(
                                                    \begin{array}{c}
                                                      r h^{(1)}_l(k r) \\
                                                      \frac{\kappa}{|\kappa|}\frac{ik r}{\varepsilon+2m_{\Lambda}}h^{(1)}_{\tilde{l}}(k r) \\
                                                    \end{array}
                                                  \right),\nonumber\\
                                     &\longrightarrow&\left(
                                                      \begin{array}{c}
                                                        1 \\
                                                        \frac{\kappa}{|\kappa|}
                                                        \frac{ik}{\varepsilon+2m_{\Lambda}} \\
                                                      \end{array}
                                                    \right)e^{ik r},
                                                    \label{Eq:behavior_rinf-1}
\end{eqnarray}
for $\varepsilon >0$ and
\begin{eqnarray}
 \phi^{(2)}_{\kappa}(r,\varepsilon)&\longrightarrow& \left(
                                                    \begin{array}{c}
                                                      r\sqrt{\frac{2Kr}{\pi}}K_{l+\frac{1}{2}}(Kr) \\
                                                     \frac{-Kr}{\varepsilon+2m_{\Lambda}} \sqrt{\frac{2Kr}{\pi}}K_{\tilde{l}+\frac{1}{2}}(Kr)\\
                                                    \end{array}
                                                  \right),\nonumber\\
                                 &\longrightarrow&\left(
                                                  \begin{array}{c}
                                                    1 \\
                                                    -\frac{K}{\varepsilon+2m_{\Lambda}}\\
                                                  \end{array}
                                                \right)e^{-Kr},
                                                \label{Eq:behavior_rinf-2}
\end{eqnarray}
for $\varepsilon<0$. Here, $K^2=(V-S-\varepsilon)(\varepsilon-V+S+2m_{\Lambda})>0$,
$h^{(1)}_l(k r)$ is the spherical Hankel function of the first kind, and
$K_{l+\frac{1}{2}}(Kr)$ is the modified spherical Bessel function.

\section{Numerical Details}
\label{Sec:Numerical}

In the present RMF-GF calculations, for the $NN$ interaction, the effective interaction PK1~\cite{PRC2004LongWH_69_034319} is taken. For the $\Lambda N$ interaction,
with $\Lambda$ hyperon mass $m_{\Lambda}=1115.6~$MeV, the scalar coupling constant $g_{\sigma\Lambda}=0.618g_{\sigma}$ is fixed to reproduce the experimental binding energies of $\Lambda$ in the $1s_{1/2}$ state of hypernucleus $^{40}_{\Lambda}$Ca ($B_{1s}^{\Lambda}=18.7~{\rm MeV}$)~\cite{PRC1999Usmani_60_055215} based on the $NN$ interaction, the vector coupling constant $g_{\omega\Lambda}=0.666g_{\omega}$ is determined from the n\"{a}ive quark model~\cite{PPNP1984Dover_12_171}, and the tensor coupling constant $f_{\omega\Lambda}=-1.0g_{\omega\Lambda}$ is taken as in Ref.~\cite{PRC1994Mares_49_2472} which is related with the spin-orbit splitting of $\Lambda$ hyperons.
With those $NN$ and $\Lambda N$ interactions, the single-$\Lambda$ binding energy $B_{\Lambda}$ for hypernuclei from $^{12}_{\Lambda}$C to $^{208}_{\Lambda}$Pb are well described and consistent results with the experimental data~\cite{PRC2001Hotchi_64_044302,PPNP2006Hashimoto_57_564} are obtained as shown in Fig.~\ref{Fig2-added}.

The RMF Dirac equation is solved in a box of size $R=20$~fm and a step size of $0.05$~fm. In the present work, single- or double-$\Lambda$ hypernuclei are studied, in which the $\Lambda$ hyperon(s) occupy(s) the $1s_{1/2}$ orbit.
To perform the integrals of the Green's function in Eq.~(\ref{EQ:density-R-GF}), the contour path $C_{\varepsilon}$ is chosen to be a rectangle with height $\gamma=0.1~$MeV and enclose only the bound state $1s_{1/2}$ on the complex energy plane as shown in Fig.~\ref{Fig1}. The energy step is taken as $d\varepsilon=0.005$~MeV on the contour path for the integral.
With these parameters of the contour path $C_{\varepsilon}$, the convergence of the obtained densities for $\Lambda$ hyperons in Eq.~(\ref{EQ:density-R-GF}) is up to $10^{-14}~{\rm fm}^{-3}$.
To calculate the density of states $n_{\kappa}(\varepsilon)$ along the real-$\varepsilon$ axis, the parameter $\epsilon$ in Eq.~(\ref{EQ:DOS-lj}) is taken as $1\times10^{-6}~{\rm MeV}$ and the energy step along the real-$\varepsilon$ axis is $1\times10^{-4}~{\rm MeV}$. With this energy step, the accuracy for energies and widths of the s.p.~resonant states can be up to $0.1~{\rm keV}$.

\section{RESULTS AND DISCUSSION}
\label{Sec:Results}

In this part, firstly, we take $^{61}_{\Lambda}{\rm Ca}$ as an example, and extend the RMF-GF model to investigate the s.p.~spectrum of hypernuclei.

In Fig.~\ref{Fig2}, the density of states $n_{\kappa}(\varepsilon)$ in different blocks $\kappa$ for the $\Lambda$ hyperon in hypernucleus $^{61}_{\Lambda}{\rm Ca}$ are plotted as a function of single-$\Lambda$ energy $\varepsilon$. The dotted line in each panel indicates the continuum threshold. The peaks of $\delta$-functional shape below the continuum threshold correspond to bound states and spectra with $\varepsilon>0$ are continuous. By comparing density of states for $^{61}_{\Lambda}$Ca (denoted by blue solid line) and those for free particles obtained with zero potential $V=S=0$ (denoted by the red solid line), one can easily find out the resonant states in the continuum. It is clear that the density of states $n_{\kappa}(\varepsilon)$ for the resonant states sit atop of those for free particles. Accordingly, the $\Lambda$ hyperon bound states are observed in $s_{1/2}$, $p_{1/2}$, $p_{3/2}$, $d_{3/2}$ and $d_{5/2}$ blocks and the resonant states are observed in $p_{1/2}$, $p_{3/2}$, $f_{5/2}$, $f_{7/2}$, $g_{7/2}$ and $g_{9/2}$ blocks.

From the density of states, we can extract the energies for the $\Lambda$ hyperon bound states and the energies ($\varepsilon_{\rm res.}$) and widths ($\Gamma$) for the resonant states. Here, $\varepsilon_{\rm res.}$ and $\Gamma$ are defined as the positions and the FWHM of resonant peaks, which are the differences between the density of states for the $\Lambda$ hyperon in $^{61}_{\Lambda}$Ca and free hyperon. We list in part (a) of Table~\ref{Tab1} the s.p.~energy $\varepsilon$ for bound states, in comparison with those obtained by the shooting method with box boundary condition, and in part (b) the energies $\varepsilon_{\rm res.}$ and widths $\Gamma$ of resonant states. From Table~\ref{Tab1}, it can be seen that s.p.~energies for bound states obtained by the Green's function method and shooting method are equal.
Six resonant states with very different widths are obtained. Very close to the continuum threshold, resonant states $2p_{1/2}$ and $2p_{3/2}$ with width $\Gamma\sim 0.1$~MeV are observed; at slightly higher energy around $0.6-0.8$~MeV, very narrow resonant states $1f_{5/2}$ and $1f_{7/2}$ with $\Gamma\sim 0.02$~MeV are observed, the behavior of these narrow resonant states is similar as bound states; and at very high energy region, much wider resonant states $1g_{9/2}$ and $1g_{7/2}$ with $\Gamma>1.1$~MeV are observed, their properties is similar as nonresonant scattering states.

\begin{table}[t!]
\center
\caption{Single-$\Lambda$ energies in $^{61}_{\Lambda}{\rm Ca}$ extracted from $n_{\kappa}(\varepsilon)$ in Fig.~\ref{Fig2} by RMF-GF method. Part (a) is for the bound states, in comparison with energies $\varepsilon_{\rm box}$ obtained by the shooting method with the box boundary condition; Part (b) is for the resonant states, where both the energies $\varepsilon_{\rm res.}$ and widths $\Gamma$ are listed. All quantities are in MeV.}
\begin{tabular}{ccccccc}
  \hline\hline
 (a)  & $ 1s_{1/2}$ & $1p_{3/2}$ & $1p_{1/2}$ & $1d_{5/2}$ & $1d_{3/2}$ & $2s_{1/2}$ \\ \hline
$\varepsilon_{\rm GF}$   & $-20.6035$  & $ -13.3109$& $-13.1363$ & $-6.0358$  & $-5.7893$  & $-5.0977$\\
$\varepsilon_{\rm box}$  & $-20.6035$  & $-13.3109$ & $-13.1363$ & $-6.0358$  & $-5.7893$  & $-5.0977$\\
\hline\hline
 (b)  &$2p_{3/2}$ &$2p_{1/2}$ & $1f_{7/2}$ & $1f_{5/2}$ & $1g_{9/2}$ & $1g_{7/2}$\\\hline
$\varepsilon_{\rm res.}$ &0.0774&0.1050&0.6147&0.8215&6.8017&6.9772\\
$\Gamma$ &0.1015&0.1259&0.0124&0.0229&3.2003&3.2926\\
\hline\hline
\end{tabular}
\label{Tab1}
\end{table}

\begin{figure}[!ht]
 \includegraphics[width=0.4\textwidth]{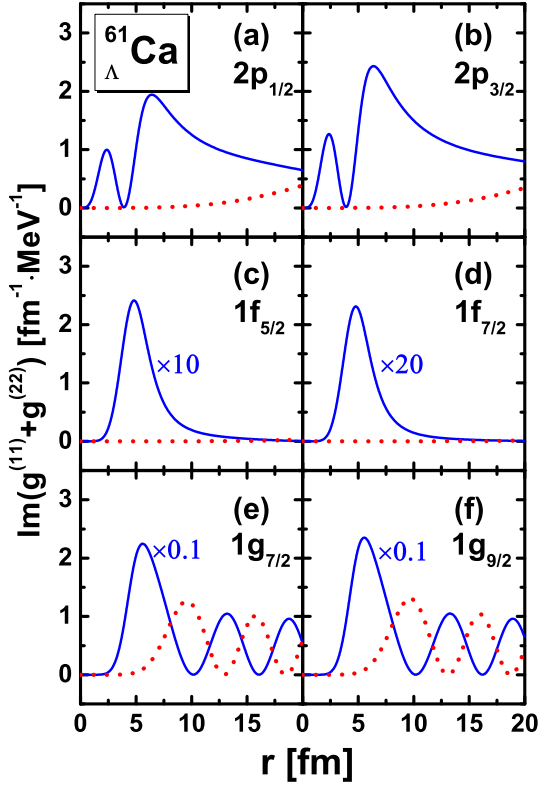}
 \caption{(Color online) The integrands for the density of states,
 ${\rm Im}[\mathcal{G}^{(11)}_{\kappa}(r,r;\varepsilon+i\epsilon)
 +\mathcal{G}^{(22)}_{\kappa}(r,r;\varepsilon +i\epsilon)]$, in Eq.~(\ref{EQ:DOS-lj})~at resonant energy $\varepsilon=\varepsilon_{\rm res.}$ for the single-$\Lambda$ resonant states $2p_{1/2}$~(a), $2p_{3/2}$~(b), $1f_{5/2}$~(c), $1f_{7/2}$~(d), $1g_{7/2}$~(e) and $1g_{9/2}$~(f) in $^{61}_{\Lambda}{\rm Ca}$ (blue solid line), in comparison with those for the free particles obtained with $V=S=0$ (red dotted line). The values of $\varepsilon_{\rm res.}$ are listed in Table~\ref{Tab1}. The integrands for $1f_{5/2}$, $1f_{7/2}$, $1g_{7/2}$ and $1g_{9/2}$ are divided by a factor of 10, 20, 0.1 and 0.1, respectively.}
 \label{Fig3}
\end{figure}

To see the distributions of the $\Lambda$ hyperon resonant states given in Table~\ref{Tab1}, we show in Fig.~\ref{Fig3} the integrands for the density of states $n_{\kappa}(\varepsilon)$, i.e., ${\rm Im}[\mathcal{G}_{\kappa}^{(11)}(r,r;\varepsilon+i\epsilon)+\mathcal{G}_{\kappa}^{(22)}(r,r;\varepsilon+i\epsilon)]$, in Eq.~(\ref{EQ:DOS-lj}) at the resonant energies. The integrand ${\rm Im}[\mathcal{G}_{\kappa}^{(11)}(r,r;\varepsilon+i\epsilon)+\mathcal{G}_{\kappa}^{(22)}(r,r;\varepsilon+i\epsilon)]$, which is calculated from the s.p.~wave functions with Eq.~(\ref{EQ:GF-RDirac}), corresponds to the particle density $\rho_{V\Lambda}$ of Eq.~(\ref{Eq:Rrhov}) at energy $\varepsilon$. From Fig.~\ref{Fig3}, it can be seen that the integrands of the resonant states with the same angular momentum $l$ have very similar distributions and very different for those with different $l$. The distributions of the resonant states are tightly related with their widths. For the $2p$ resonant states, the distributions at resonant energies are very extended and have large components at coordinate space with $r > 5~{\rm fm}$. On the contrary, for the very narrow $1f_{5/2}$ and $1f_{7/2}$ resonant states, the density at resonant energy mainly localized around the surface, i.e., $2.5~{\rm fm} < r < 7.5~{\rm fm}$ with a maximum around $r = 5~{\rm fm}$, the behaviors are very similar as bound state; and for the very wide $1g_{7/2}$ and $1g_{9/2}$ resonant states, the distribution is scattering and outgoing, the behaviors are very similar as the nonresonant scattering states shown by the red dotted lines.

\begin{table}[!h]
\center
\caption{Comparison of the energies $\varepsilon_{\rm res.}$ and widths $\Gamma$ of the single-neutron resonant states in (hyper)nuclei $^{60}{\rm Ca}$, $^{61}_{\Lambda}{\rm Ca}$ and $^{62}_{2\Lambda}{\rm Ca}$ obtained by RMF-GF method. All quantities are in MeV.}
\begin{tabular}{cccccr}
  \hline\hline
  &&~$1g_{9/2}$~&~$2d_{5/2}$~&~$1g_{7/2}$~&~$1h_{11/2}$\\\hline
\multirow{2}*{$^{60}{\rm Ca}$} &
$\varepsilon_{\rm res.}$&0.7656&1.0722&5.4906&10.4430\\
               &$\Gamma$&0.0012&0.4134&0.8710&1.9785\\\hline
\multirow{2}*{$^{61}_{\Lambda}{\rm Ca}$}&
 $\varepsilon_{\rm res.}$ &0.6679&1.0497&5.4978&10.3981\\
                 &$\Gamma$&0.0009&0.3915&0.8746&1.9710\\\hline
\multirow{2}*{$^{62}_{2\Lambda}{\rm Ca}$}&
$\varepsilon_{\rm res.}$ &0.5703&1.0260&5.5049&10.3526\\
                &$\Gamma$&0.0007&0.3712&0.8795&1.9644\\
  \hline\hline
\end{tabular}
\label{Tab2}
\end{table}
\begin{figure}[!t]
 \includegraphics[width=0.4\textwidth]{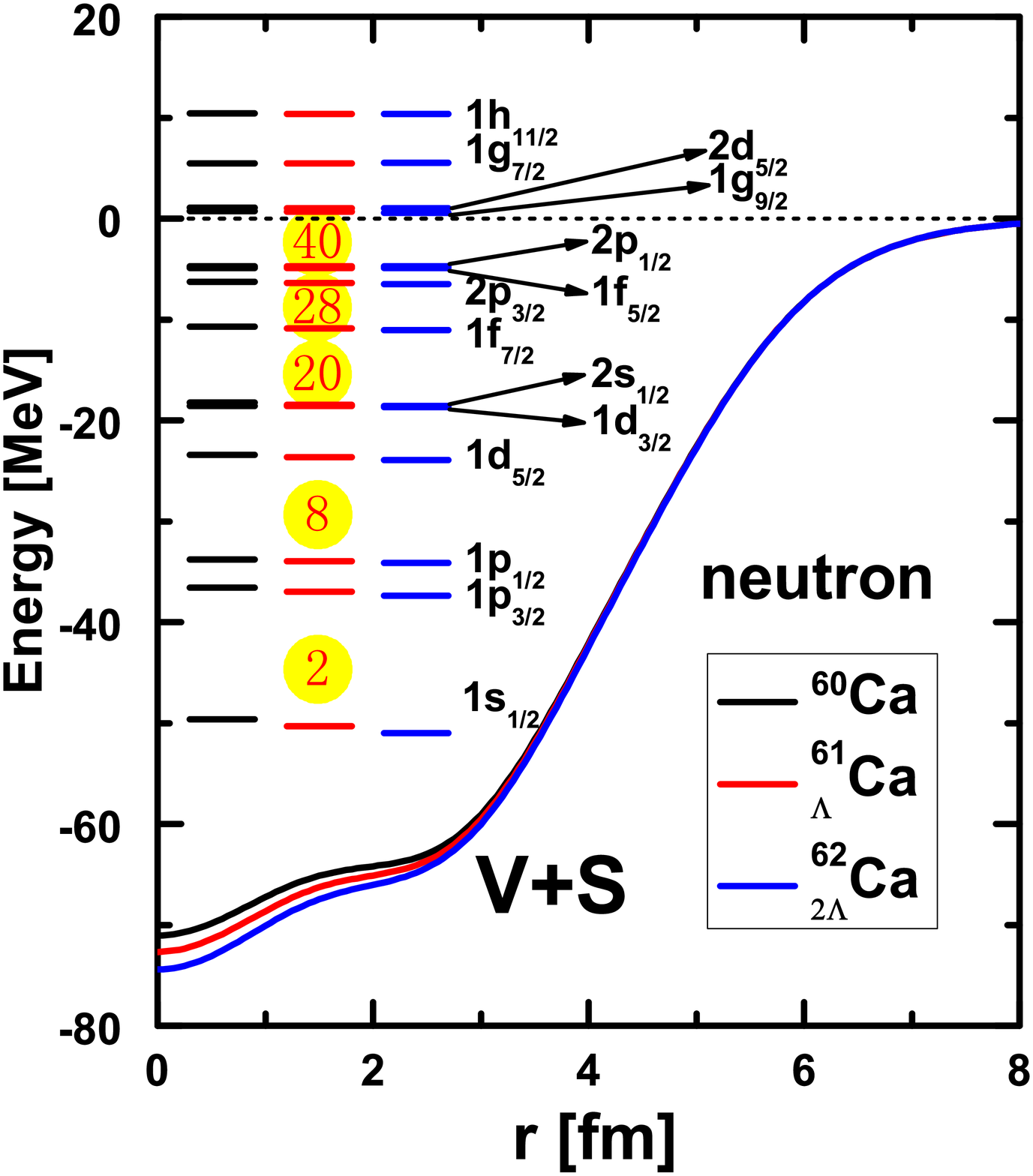}
 \caption{(Color online) Mean field potentials $V(r)+S(r)$ as well as the s.p.~levels for neutrons in (hyper)nuclei $^{60}{\rm Ca}$, $^{61}_{\Lambda}{\rm Ca}$ and $^{62}_{2\Lambda}{\rm Ca}$ obtained by RMF-GF method.}
 \label{Fig4}
\end{figure}

\begin{figure}[!t]
 \includegraphics[width=0.4\textwidth]{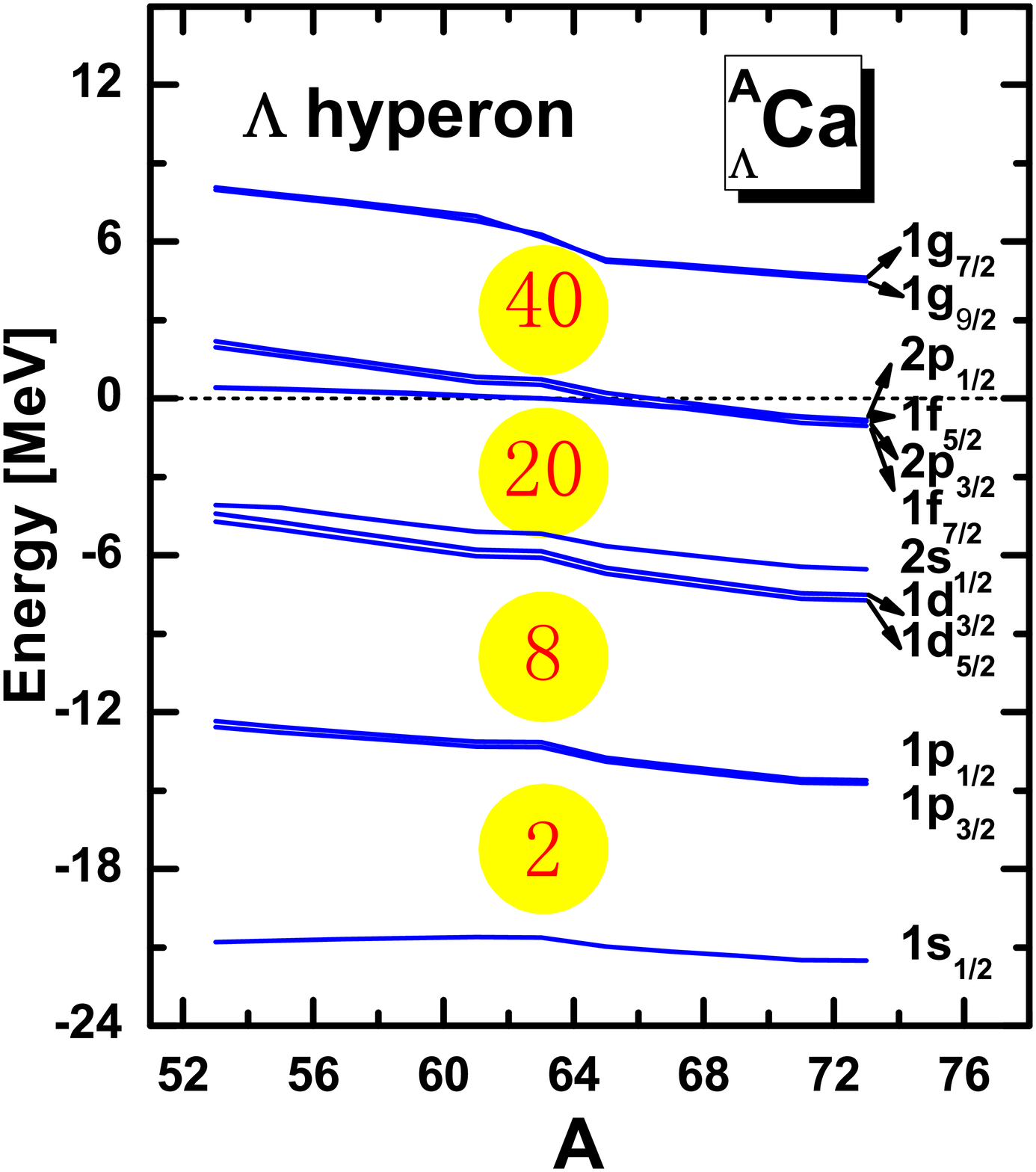}
 \caption{(Color online) Evolution of the s.p.~levels for the $\Lambda$ hyperon in the $\rm{Ca}$ isotopes as a function of mass number $A$ calculated by RMF-GF method.}
 \label{Fig5}
\end{figure}

It is well known that neutron, proton and $\Lambda$ hyperon obey their own Pauli Principle since they are different Fermions.
However, in the self-consistent RMF model, $\Lambda$ hyperon is glue-like and will influence the properties of nucleons.
In this part, taking $^{60}$Ca, $^{61}_{\Lambda}{\rm Ca}$ and $^{62}_{2\Lambda}{\rm Ca}$ as examples, we investigate the influences of $\Lambda$ hyperons on the single-neutron resonant states.
In Table~\ref{Tab2}, the energies $\varepsilon_{\rm res.}$ and widths $\Gamma$ of the single-neutron resonant states in these (hyper)nuclei obtained by RMF-GF method are listed. Four single-neutron resonant states $1g_{9/2}$, $2d_{5/2}$, $1g_{7/2}$, and $1h_{11/2}$ are obtained, and their energies $\varepsilon_{\rm res.}$ and widths $\Gamma$ decrease with the increase of the number of $\Lambda$ hyperon except orbit $1g_{7/2}$ which increases slightly.

To investigate the changes of the single-neutron resonant states brought by adding $\Lambda$ hyperons to $^{60}$Ca shown in Table~\ref{Tab2}, the mean-field potential $V+S$ as well as the s.p.~levels including the bound states and resonant states for neutrons in (hyper)nuclei $^{60}$Ca, $^{61}_{\Lambda}$Ca and $^{62}_{2\Lambda}$Ca are plotted in Fig.~\ref{Fig4}. Adding more $\Lambda$ hyperons make the central part of the neutron mean-field potential become around $1.5~$MeV depressed per hyperon due to the attractive $\Lambda N$ interaction. As a result, the s.p.~levels for neutrons go down with the increase of the number of $\Lambda$ hyperons.

Finally, the energy level structures for $\Lambda$ hyperons are studies. In Fig.~\ref{Fig5}, we plot the single-$\Lambda$ energies $\varepsilon_{\Lambda}$ for the Ca hypernucleus isotopes as a function of the mass number $A$. It can be seen that with increasing hypernuclei mass, the s.p.~levels for $\Lambda$ hyperon go down. Obvious shell gaps are found for $\Lambda$ hyperon s.p.~levels. Besides, the spin-orbit splitting between the spin doublet states $1p$, $1d$, $1f$ and $1g$ are much smaller than those for nucleons shown in Fig.~\ref{Fig4}.
Experimentally, the spin-orbit splitting between the $1p_{1/2}$ and the $1p_{3/2}$ hyperon states in $^{13}_{\Lambda}{\rm C}$ was found to be much smaller than the spin-orbit splitting in ordinary nuclei by a factor of $20-30$~\cite{PRL2001Ajimura_86_4255}. Other experiments~\cite{PLB1978WBruckner_79_157} got the same conclusion. Our present results are consistent with those experimental data.
In Fig.~\ref{Fig5}, low lying $2p$ orbits are found in the continuum, which play important roles in forming hyperon halos.
In Ref.~\cite{CPL2002HFLu_19_1775}, the hyperon halo in $^{15}_{3\Lambda}$C and $^{16}_{4\Lambda}$C is predicted by the relativistic continuum Hartree-Bogoliubov (RCHB) theory and due to the occupation of the weakly bound state $1p_{3/2}$ with
extended density distributions and small separation energy of the $\Lambda$ hyperons.
According to those studies, we prefer to say hyperon halo may appear in the Ca hypernucleus isotopes due to the low lying or weakly bound $2p$ orbits.
\section{SUMMARY}\label{Sec:summary}

In this work, the RMF theory with Green's function method in coordinate space is extended to investigate $\Lambda$ hypernuclei. Detailed formula are presented.

Firstly, taking $^{61}_{\Lambda}$Ca as an example, the RMF-GF model is applied to study the single-$\Lambda$ resonant states. By analyzing the density of states, the s.p.~energy for bound states and energies and widths for the resonant states are obtained. Consistent results for the single-$\Lambda$ bound states between the Green's function method and shooting method are obtained. Six resonant states are observed with very different widths, and the distributions of the very narrow $1f_{5/2}$ and $1f_{7/2}$ states are very similar as bound states while the distributions of the wide $1g_{7/2}$ and $1g_{9/2}$ states are like scattering states.

Secondly, taking $^{60}$Ca, $^{61}_{\Lambda}$Ca and $^{62}_{2\Lambda}$Ca as examples, we investigate the influences of $\Lambda$ hyprons on the single-neutron resonant states and found that for most resonant states, with the increase of the number of $\Lambda$ hyperon, both the energies and widths decrease due to the deeper mean-field potential.

Finally, the s.p.~level for $\Lambda$ hyperon in the Ca isotopes are studied. Obvious shell structure is found for $\Lambda$ hyperon and very small spin-orbit splitting is obtained, which is consistent with the present experimental results.

\begin{acknowledgments}
This work was partly supported by the National Natural Science Foundation of China (Grant Nos.~11175002, 11335002, 11505157, 11675148, and 11105042).
\end{acknowledgments}


\end{document}